\documentstyle[12pt]{article}
\begin{document}
\newcommand{\eq}{\begin{eqnarray}}
\newcommand{\en}{\end{eqnarray}}
\begin{center}
{\Large\bf
Perturbative framework for the $\pi^+\pi^-$ atom
}
\end{center}
\begin{center}
{\bf
V. E. Lyubovitskij$^{1,2,{\dag}}$,
E. Z. Lipartia$^{3,4,{\#}}$ and
A. G. Rusetsky$^{1,4,{\ddag}}$
}
\end{center}
{\em
\noindent
$^1$Bogoliubov Laboratory of Theoretical Physics, JINR, 141980 Dubna, Russia

\noindent
$^2$Department of Physics, Tomsk State University, 634050 Tomsk, Russia

\noindent
$^3$Laboratory for Computational Technique and Automation, JINR, 141980 Dubna, Russia

\noindent
$^4$IHEP, Tbilisi State University, 380086 Tbilisi, Georgia

\noindent
$^{\dag}$e-mail: LUBOVIT@THSUN1.JINR.DUBNA.SU

\noindent
$^{\#}$e-mail: LIPARTIA@THSUN1.JINR.DUBNA.SU

\noindent
$^{\ddag}$e-mail: RUSETSKY@THSUN1.JINR.DUBNA.SU
}
\begin{abstract}
The perturbative framework is developed for the calculation of the
$\pi^+\pi^-$ atom characteristics on the basis of the field-theoretical
Bethe-Salpeter approach.
A closed expression for the first-order correction to the $\pi^+\pi^-$
atom lifetime has been obtained.
\end{abstract}


\vspace*{.6cm}
At present time the experiments on the study of hadronic atoms
$\pi\pi$ \cite{NM}, $\pi p$, $\pi d$ \cite{G} have
being carried out. Namely, the first estimate of the $\pi^+\pi^-$  atom
lifetime was given in Ref.~\cite{NM}. Now the DIRAC
collaboration works out an experiment at CERN on the high precision
measurement of the lifetime of $\pi^+\pi^-$ atoms. This experiment might
provide a decisive improvement in the direct determination of the
difference of the $S$-wave $\pi\pi$ scattering lengths and
thus serve as a valuable test for the
predictions of Chiral Perturbation Theory~\cite{CH}.
In the view of these experiments there
arises a need in the theoretical framework which would enable one to
calculate the characteristics of such atoms with a high accuracy based
on the ideas of standard model.

The theoretical study of hadronic atoms starts from
Refs.~\cite{DE}-\cite{BL} where the nonrelativistic relations
of the energy level displacement of the hadronic atom due to strong
interactions and its lifetime with the strong scattering lengths are
established.
The expression for the width $\Gamma_0$ of the $\pi^+\pi^-$ atom in
the ground state is
\eq\label{fd}
\Gamma_0\,\,=\,\,\frac{16\pi}{9}\,\,\sqrt{\frac{2\Delta m_\pi}{m_\pi}}
\,\,(a^0_0-a^2_0)^2\,\,\phi_0^2
\en
\noindent where $\Delta m_\pi$ is the $m_{\pi^\pm}-m_{\pi^0}$ mass
difference, and $\phi_0$ is the value of the Coulomb wave function
(w.f.) of the pionium at the origin.

The approach to the study of the problem of hadronic atoms, developed in
Ref.~\cite{DE}, makes use of the general characteristic feature of
the hadronic atoms~-- the factorization of strong and electromagnetic
interactions. The formula (\ref{fd})
demonstrates this factorization property explicitly, expressing the atom
lifetime as a product of two factors~-- the Coulomb w.f. at the
origin and the strong interaction factor, completely concentrated in
the $\pi\pi$ strong scattering lengths.

The problem of evaluation of the electromagnetic and strong corrections
to the basic formula (\ref{fd}) within different approaches is
addressed in Refs.~\cite{TR}-\cite{SZ}. For a brief review
see Ref.~\cite{LR}. In this paper within the Bethe-Salpeter (BS)
approach we have derived the relativistic analogue of the formula
(\ref{fd}) taking into account the correction due to the displacement
of the bound state pole position by strong interactions (strong correction)
in the first order. This correction was found to be of the relative
order $10^{-3}$. It should be stressed that the field-theoretical
approaches \cite{SL,LR,KV,SZ} to the problem, unlike the
potential treatment~\cite{TR,RS}, do not refer to a concept
of the phenomenological strong interaction $\pi\pi$ potential, which is
a source of an additional ambiguity in the calculations of hadronic atom
characteristics. In the former approaches these characteristics are
expressed directly in terms of the underlying strong interaction (chiral)
Lagrangian, and the results can be compared to the experiment, providing
the consistent test of the predictions of chiral theory.

In the present work we suggest a relativistic perturbative framework
for the calculation of the energy levels and lifetime of hadronic atoms.
The main purpose of this work is to
demonstrate a possibility (not only in the potential scattering theory,
but in the BS treatment as well) of the clear-cut factorization of strong
and electromagnetic interactions in the observable characteristics of
hadronic atoms, avoiding the double-counting problem in the calculation
of these quantities. One should note that the suggested approach
allows to calculate strong and electromagnetic corrections in all orders
of the perturbation theory. At the present stage we apply the general
formalism to the calculation of the first-order strong and electromagnetic
corrections to the pionium lifetime. The results for strong corrections
obtained in Ref.~\cite{LR} are reproduced in these calculations.

Our framework is based on the perturbative expansion which is performed
around the solution of the BS equation with the Coulomb kernel similar
to that introduced in Ref.~\cite{BR}
\eq\label{vc}
V_C({\bf p},{\bf q})=\sqrt{w({\bf p})}
\frac{4im_\pi e^2}{({\bf p}-{\bf q})^2}\sqrt{w({\bf q})}, \hspace*{1.5cm}
w({\bf p})=\sqrt{m^2_\pi+{\bf p}^2}
\en
\noindent
The factor $\sqrt{w({\bf p})w({\bf q})}$ introduced in the kernel (\ref{vc})
enables one to reduce the BS equation with such a kernel to the exactly
solvable Schr\"odinger equation with the Coulomb potential.
Then, the exact solution of the BS equation
with this kernel is written in the form
\eq\label{wfc}
\psi_C(p)=iG_0(M^\star;p)\,4\sqrt{w({\bf p})}\,\,
\frac{4\pi\alpha m_\pi\phi_0}{{\bf p}^2+\gamma^2}\, , \,\,\,\,\,
\bar\psi_C(p)=\psi_C(p),
\en
\noindent where $\gamma=m_\pi\alpha/2$ and ${M^\star}^2=m_\pi^2(4-\alpha^2)$
is the eigenvalue corresponding to the unperturbed ground-state solution.
$G_0$ denotes the free Green's function of the $\pi^+\pi^-$-pair.
The exact Green's function corresponding to the Coulomb kernel (\ref{vc})
is given by the well-known expression
\eq\label{gc4}
\hspace*{-.7cm}G_C(P^\star;p,q)&=&(2\pi)^4\delta^{(4)}(p-q)G_0(P^\star;p)+
G_0(P^\star;p)T_C(E^\star;{\bf p},{\bf q})
G_0(P^\star;q)
\en
\noindent Here $T_C$ is given by
\eq\label{gc3}
T_C(E^\star;{\bf p},{\bf q}) &=& 16 i \pi m_\pi \alpha
\sqrt{w({\bf p})w({\bf q})}\,\,\,\biggl[\frac{1}{({\bf p}-{\bf q})^2}
+ \int_0^1\frac{\nu d\rho \rho^{-\nu}}{D(\rho;{\bf p}, {\bf q})}\biggr]\\[2mm]
D(\rho;{\bf p}, {\bf q}) &=&
({\bf p}-{\bf q})^2\rho-\frac{m_\pi}{4E^\star}
\biggl(E^\star-\frac{{\bf p}^2}{m_\pi}\biggr)
\biggl(E^\star-\frac{{\bf q}^2}{m_\pi}\biggr)(1-\rho)^2,
\nonumber
\en
\noindent where $\nu=\alpha\sqrt{m_\pi/(-4E^\star)}$ and
$E^\star=({P^\star}^2-4m_\pi^2)/(4m_\pi)$.

The full BS equation for the $\pi^+\pi^-$ atom w.f. $\chi(p)$ is
written as
\eq\label{bs}
G_0^{-1}(P;p)\chi(p)=\int\frac{d^4k}{(2\pi)^4}V(P;p,q)\chi(q),
\en
\noindent where $V(P;p,q)$ denotes the full BS kernel which is
constructed from the underlying (effective) Lagrangian according
to the general rules and includes all strong and electromagnetic
two-charged-pion irreducible diagrams. In particular, it contains
the diagrams with two neutral pions in the intermediate state which
determine the decay the $\pi^+\pi^-$ atom into $\pi^0\pi^0$. Note
that in addition $V(P;p,q)$ contains the charged pion self-energy
diagrams attached to the outgoing pionic legs (with the relative
momentum $q$), which are two-particle reducible. These diagrams
arise in the definition of the kernel $V(P;p,q)$ because the free
two-particle Green's function is used in the l.h.s. of Eq. (\ref{bs})
instead of the dressed one.
The c.m. momentum squared $P^2$ of the atom has the complex value,
corresponding to the fact that the atom is an unstable system.
According to the conventional parametrization, we can write
$P^2=\bar M^2=M^2-iM\Gamma$ where $M$ denotes the "mass" of the atom,
and $\Gamma$ is the atom decay width.

The full four-point Green's function $G(P)$ for the kernel $V$ has a pole
in the complex $P^2$ plane at the bound-state energy. The relation between
the exact w.f. $\chi(p)$ and the Coulomb w.f.
$\psi_C$ is given by~\cite{LR}
\eq\label{in}
<\chi|=C<\psi_C|\, G_C^{-1}(P^\star)G(P),\,\,\,\,\,
{{P^\star}^2\to{M^\star}^2,~P^2\to\bar M^2}
\en
\noindent where $C$ is the normalization constant.
In what follows we assume that the limiting procedure is performed with
the use of the prescription~\cite{LR}
${P^\star}^2={M^\star}^2+\lambda,\,P^2=\bar M^2+\lambda,\,\lambda\to 0$.
The validity of Eq. (\ref{in}) can be trivially checked, extracting
the bound-state pole in $G(P)$ and using the BS equation for $\psi_C$.

In order to perform the perturbative expansion of the bound-state
characteristics $M$ and $\Gamma$ around the unperturbed values we,
as in Ref.~\cite{LR}, split the full BS kernel $V$ into two parts as
$V=V_C+V'$ and consider $V'$ as a perturbation. It can be shown that
Eq. (\ref{in}) is equivalent to
\eq\label{newwf}
<\chi|=-C^{-1}<\psi_C|\bigl[1+(\Delta G_0^{-1}-V')G_RQ\bigr]^{-1},
\hspace*{1.5cm}
\Delta G_0^{-1}=G_0^{-1}(P)-G_0^{-1}(P^\star)
\en

With the use of Eq. (\ref{newwf}) the following identity is easily obtained
\eq\label{basic}
<\psi_C|\bigl[1+(\Delta G_0^{-1}-V')G_RQ\bigr]^{-1}
(\Delta G_0^{-1}-V')|\psi_C>=0,
\en
\noindent which is an exact relation and serves as a basic equation
for performing the perturbative expansion for the bound-state energy.

In the Eqs. (\ref{newwf}) and (\ref{basic}) $G_RQ$ stands for the
regular (pole subtracted) part of the Coulomb Green's function
(\ref{gc4}), projected onto the subspace, orthogonal to the ground-state
unperturbed solution. This quantity can be further split into two pieces,
according to $G_RQ=G_0(M^\star)+\delta G$. Here the function $\delta G$
corresponds to the ladder of the exchanged Coulomb photons and thereby
contains explicit powers of $\alpha$. It is given by the following expression:
\eq\label{deltag}
\hspace*{-0.8cm}\delta G &=&
i\sqrt{w({\bf p})w({\bf q})}\biggl[\Phi({\bf p}, {\bf q})
- S({\bf p})S({\bf q})
\frac{8}{M^\star}
\frac{\partial}{\partial M^\star}\biggr]
G_0(M^\star,p)G_0(M^\star,q)\nonumber\\
\hspace*{-0.8cm}\Phi({\bf p}, {\bf q}) &=& 16\pi m_\pi\alpha
\biggl[\frac{1}{({\bf p} - {\bf q})^2} + I_R({\bf p}, {\bf q})\biggr]
+ (m_\pi\alpha)^{-2} S({\bf p})S({\bf q}) R({\bf p}, {\bf q})\\
\hspace*{-0.8cm}S({\bf p}) &=&
4\pi m_\pi\alpha\phi_0({\bf p}^2 + \gamma^2)^{-1},\,\,
R({\bf p}, {\bf q}) = 25 - \sqrt{\frac{8}{\pi m_\pi\alpha}}
[S({\bf p}) + S({\bf q})]+\cdots\nonumber
\en
\noindent where the ellipses stand for the higher-order terms in $\alpha$.
The integral $I_R({\bf p}, {\bf q})$ is given by
\eq
I_R({\bf p}, {\bf q}) = \int\limits^1_0 \frac{d\rho}{\rho}
\,\,\,[D^{-1}(\rho;{\bf p}, {\bf q}) - D^{-1}(0;{\bf p}, {\bf q})],\,\,\,
E^\star=-\frac{1}{4}m_\pi\alpha^2
\en

The equation (\ref{newwf}) expresses the exact BS w.f. of the atom in
terms of the unperturbed w.f. via the perturbative expansion in the
perturbation potential $V'$. This potential consists of the
following pieces:\\
\hspace*{.5cm}
1. The purely strong part, which is isotopically invariant.
This part survives when the electromagnetic interactions are
"turned off" in the Lagrangian.\\
\hspace*{.5cm}
2. The part, containing the diagrams with the finite mass
insertions which are responsible for the $m_{\pi^\pm}-m_{\pi^0}$
electromagnetic mass difference.\\
\hspace*{.5cm}
3. The part, containing the exchanges of one, two, ... virtual
photons and an arbitrary number of strong interaction vertices.

Note that the terms 1 and 2 are more important due to the following
reasons. The first term includes strong interactions
responsible for the decay of the pionium. The second term makes
this decay kinematically allowed due to finite difference of
charged and neutral pion masses. Consequently, it seems to be natural
to consider together the pieces 1 and 2. We refer to the corresponding
potential as $V_{12}$.
The $T$-matrix corresponding to the potential $V_{12}$ is defined by
$T_{12}(P)=V_{12}(P)+V_{12}(P)G_0(P)T_{12}(P)$.
The rest of the potential $V'$ is referred as $V_{3}=V'-V_{12}$.
In what follows we restrict ourselves to the first order in the fine
structure constant $\alpha$, i.e. consider the diagrams with only one
virtual photon contained in $V_3$.

Returning to the basic equation (\ref{basic}), we expand it in the
perturbative series considering $V_3$ and $\delta G$ as perturbations.
Meanwhile we expand $\Delta G_0^{-1}$ in the Taylor series in
$\delta M=\bar M-M^\star$ and substitute
$\bar M=M^\star+\Delta E^{(1)}+\Delta E^{(2)}-{i}/{2}\,\Gamma^{(1)}-{i}/{2}
\,\Gamma^{(2)}+{(8M^\star)}^{-1}{{\Gamma^{(1)}}^2}+\cdots$.

Restricting
ourselves to the first order of the perturbative expansion, we arrive at
the following relations
\eq\label{or}
\Delta E^{(1)}={\rm Re}\left(\frac{i}{2M^\star}
\frac{T_{12}}{m_\pi}\phi_0^2\right),\,\,\,\,
-\frac{1}{2}\Gamma^{(1)}={\rm Im}\left(\frac{i}{2M^\star}
\frac{T_{12}}{m_\pi}\phi_0^2\right)
\en
\noindent
Hereafter we use the local approximation for $T_{12}$, assuming that it does
not depend on the relative momenta. The Eqs. (\ref{or}) coincide
with the well-known Deser-type formulae for the energy-level displacement
and lifetime~\cite{DE}. Note that on the mass shell
\eq
\hspace*{-.7cm}
{\rm Re}(iT_{12})\sim T(\pi^+\pi^-\to\pi^+\pi^-),\,\,\,
{\rm Im}(iT_{12})\sim \sqrt{\Delta m_\pi}
|T(\pi^+\pi^-\to\pi^0\pi^0)|^2
\en
If we assume $V_3=\delta G=0$, we arrive at the result
\eq\label{strong}
\frac{\Gamma^{(2)}}{\Gamma^{(1)}}=
-\frac{9}{8}\frac{\Delta E^{(1)}}{E_1}-0.763\alpha,
\hspace*{1.5cm}E_1=-\frac{1}{4}m_\pi\alpha^2
\en
The first term of this expression called "strong correction"
was obtained in our previous paper~\cite{LR}. However,
as opposed to the present derivation, in Ref.~\cite{LR}
we have used the Born approximation for the calculation of
$\Delta E^{(1)}$, i.e. in Eq. (\ref{or}) $T_{12}$ was
substituted by $V_{12}$. The last term comes from the relativistic
normalization factor $\sqrt{w({\bf p})w({\bf q})}$ in the
kernel~(\ref{vc}) and corresponds to relativistic modification of the
pionium Coulomb w.f.
$|\int d^4p/(2\pi)^4\psi_C(p)|^2=\phi_0^2(1-0.381\alpha)^2/m_\pi$.
Since this correction comes from the Coulomb w.f. of the atom, it
does not depend on the parameters of the strong $\pi\pi$ interaction,
and for this reason it was neglected in Ref.~\cite{LR}.

Inclusion of $\delta G$ introduces the correction due to the exchange of
the infinite number of Coulomb photons in the lifetime.
The integrals emerging in the calculation of this correction
are ultraviolet convergent, containing, however (in a complete analogy
with a well-known result from nonrelativistic scattering theory),
an infrared enhancement $\alpha{\rm ln}\alpha$ which stems from the
one-photon exchange piece in Eq. (\ref{gc3}). Collecting all terms
together and using Eqs. (\ref{or}) for relating ${\rm Im}T_{12}$ to
$\Delta E^{(1)}$, we finally arrive at the first-order correction to
the pionium rate
\eq\label{fn}
\hspace*{-.7cm}\Gamma&=&\Gamma_0\biggl(1+
\underbrace{\biggl(-\frac{9}{8}\,\frac{\Delta E^{(1)}}{E_1}\biggr)}_{\rm strong}\,\,\,
+\underbrace{(-0.763\alpha)}_{\rm relativistic~w.f.}\, + \,\,\,\,
\underbrace{\left({1}/{2}+2.694-{\rm ln}\alpha\right)
\frac{\Delta E^{(1)}}{E_1}}_{\rm Coulomb~photon~exchanges}\,\,\,
+\nonumber\\
\hspace*{-.7cm}&+&\delta_M-{\left( M^\star\Gamma^{(1)}\right)^{-1}}
{\rm Re}<\psi_C|(1+T_{12}G_0(M^\star))V_3(1+G_0(M^\star)T_{12})|\psi_C>\biggr)
\en
\noindent
\noindent
where $\delta_M$ stands for the mass shift correction \cite{SZ}
and the last term collects the radiative corrections \cite{SZ,KV}
(including retardation correction \cite{SL}, correction due to vacuum
polarization \cite{EF}, etc.). In the Eq. (\ref{fn}) all
first-order strong and electromagnetic corrections are given in
a closed form avoiding any difficulties connected with double
counting problem. The kernel which appears in the last term:
$(1+T_{12}G_0(M^\star))V_3(1+G_0(M^\star)T_{12})$,
is constructed from the underlying Lagrangian with the use
of the conventional Feynman diagrammatic technique. The detailed
reexamination of the above mentioned corrections within BS approach
will be addressed in our forthcoming publications.

In order to estimate the size of the calculated corrections to the pionium
lifetime (Eq. (\ref{fn}) we have used the following value of the singlet
scattering length $m_\pi(2a_0^0+a_0^2)=0.49$ corresponding
to the value $\Delta E^{(1)}/E_1=0.24 \%$. The first, second and third terms
then contribute, respectively, $-0.26\%$, $-0.55\%$ and $+1.85\%$,
and the total contribution amounts up to $\sim 1\%$ to the decay width
(apart from the mass shift and radiative corrections).
The largest contribution comes from the $\alpha{\rm ln}\alpha$ term
in Eq. (\ref{fn}).
\vspace*{.4cm}

We thank J. Gasser, M.A. Ivanov, E.A. Kuraev, H. Leutwyler,
P. Minkowski, L.L. Nemenov and H. Sazdjian for useful
discussions, comments and remarks. A.G.R. thanks Bern University for
the hospitality where part of this work was completed.
This work was supported in part by the INTAS Grant 94-739 and
by the Russian Foundation for Basic Research (RFBR) under contract
96-02-17435-a.

\end{document}